\documentclass[12pt]{article}
\usepackage{amssymb} 
\usepackage{amsmath}
\usepackage{makeidx}
\usepackage{graphicx}
\usepackage{geometry} 
\usepackage[parfill]{parskip}

\def\BEn{\begin{enumerate}}
\def\EEn{\end{enumerate}}

\def\o{{\circ}}
\def\fro{\kern-2pt\leftarrow\kern-2pt}

\def\Fi{\8{\imath}}

\def\0#1{{\mathrm{#1}}}
\def\1#1{{\mathbb{#1}}}
\def\2#1{{\mathbf{#1}}}
\def\3#1{{\mathcal {#1}}}
\def\4#1{{\mathsf{#1}}}
\def\5#1{{\mathfrak{#1}}}
\def\6#1{\overline{#1}}
\def\7#1{{\check{#1}}}
\def\8#1{{\widehat{#1}}}

\def\dag{\dagger}
\def\adj{{^{\dag}}}

\def\Bar{\,{\vrule height 10 pt width .7pt depth 4pt}\;}
\def\<{\langle}

\def\>{\rangle}

\def\Cl{\mathop{{\mathrm {Cl}}}\nolimits} 

\def\ox{\otimes}

\def\x{\times}

\def\BE{\begin{equation}}
\def\EE{\end{equation}}
\def\BEA{\begin{eqnarray}}
\def\EEA{\end{eqnarray}}

\newtheorem{principle}{Principle}
\newtheorem{definition}[principle]{Definition}

\begin{document}

\title%
{\huge General quantization
}
\author%
 { 
 David Ritz Finkelstein%
\thanks
{ School of Physics, 
Georgia Institute of Technology, Atlanta, Georgia. df4@mail.gatech.edu.
This updates part of a talk given at Glafka 2005, Athens.
quant-ph/0601002}
}
\maketitle

\abstract{
Segal's hypothesis that physical theories drift toward simple groups
follows from a general quantum principle 
and suggests a general quantization process.
I general-quantize the scalar meson field in Minkowski space-time
to illustrate the process.
The result is a finite quantum field theory
over a quantum space-time with higher symmetry than the singular  theory. 
Multiple quantification connects the
 levels of the theory.
}

\section{Quantization as regularization}
\label{sec:REGULARIZATION}

Quantum theory began with {\em ad hoc} 
regularization  prescriptions of Planck and Bohr
to fit the weird behavior
of the
electromagnetic field 
and  the nuclear atom
and to handle infinities
that blocked
 earlier theories. 
In 1924 Heisenberg discovered  
that one small  change in algebra
did both naturally.
In the early 1930's he suggested extending his algebraic 
method 
to space-time, to regularize field theory,
inspiring  the 
pioneering quantum space-time of Snyder \cite{WESS2002}.
Dirac's historic quantization program for gravity also
eliminated absolute space-time points from the quantum theory 
of gravity, leading
Bergmann too to say
that  
the world point itself possesses no physical reality
\cite{BERGMANN1972, BERGMANN1979}.

For many the infinities that still haunt physics
cry for further and deeper quantization,
but there has been little agreement on exactly
what and how far to quantize.
According to Segal canonical quantization  
continued a drift of physical theory toward simple groups
 that special
relativization began. 
He
proposed on Darwinian
grounds
 that  further
quantization should lead to  simple groups \cite{SEGAL}.
Vilela Mendes initiated the work in that direction
\cite{ VILELA1994}\/.

Each non-simplicity 
of the operational algebra
arises from an 
idol of the theory
 in the sense of Bacon  \cite{BACON}.
 An idol is
a false absolute
in which we believe
beyond the experimental evidence,
a construct that we assume to 
act but not to react.
Today it may be more practical to topple these idols than to 
work around them.

 Invariant subgroups  and
 Lie algebra ideals correspond to idols and
force us to infinite-dimensional representations.
Simple Lie algebras have enough 
finite-dimensional representations.
We relativize the absolute and regularize the theory
by simplifying  the Lie algebra:
slightly changing
its structural tensor 
so that the Lie algebra becomes 
simple, or at least simpler.
An arbitrarily small  homotopy 
of the structure tensor often suffices for simplicity.
Lie algebra simplification is  a key step in  special and general relativization, 
canonical quantization
and general quantization.

Physics has several levels and canonical quantization
simplifies only the highest-level Lie algebra,
and that not all the way to simplicity and finiteness.
General quantization 
extrapolates canonical quantization
in both respects.
It simplifies the Lie algebras on all the known levels of a physical theory,
and it simplifies them all the way to simplicity and finiteness.
It does this by small changes in the
structure tensor so that hopefully
it makes only small changes in experimental predictions
near
 the group identity,
 in
 the  correspondence domain.
 
For exercise and illustration
we general-quantize the scalar meson quantum field here.
A first-level quantization of space-time or the ether
resolves it
into an aggregate of
many identical finite quantum elements,
chronons.
A second-level general quantization of the field
resolves the field history into 
aggregates of chronons.
The vacuum  mode-vector $|\0{vac}\>$\/
represents the ambient mode of the ether.
General quantization infers candidate
structures and symmetries for the ether and its elements
from the structure and symmetry of the present-day vacuum
by a routine 
heuristic procedure based on correspondence, 
simplicity, and symmetry.

\section{Less is different too}

More is different \cite{ANDERSON}; different from less, one understands.
When we pass from few systems to many
we encounter self-organization, with
more structure, less symmetry.

It follows that less is different too;
different from more.
When we pass from many systems to few,
as from the ether to the sub-ether, we expect
to
lose organization, gain symmetry,
and lose structure.
General quantization
accords with this expectation.
Discretization, however,
and
bottom-up reconstructions of the sub-ether
like vortex, network, string, and loop models,
enrich its structure, reduce its symmetry,  
and increase its singularity.

Simple Lie algebras have quite special dimensions.
There is no simple Lie algebra  of dimension 2, for example, and
yet  there is the stable
two-dimensional compound (= non-semisimple)
 Lie algebra $W_2(q,p)$ with $qp-pq=q$.
Therefore general quantization often requires us to
introduce  new dynamical variables into the Lie algebra, 
called regulators, 
to raise its dimensionality to that of a simple Lie algebra.
Then 
we must also invoke a self-organization,
a spontaneous symmetry-breaking,
to freeze out these new variables
and recover the singular theory
in a limited correspondence domain.
Special relativization and canonical quantization
introduced no regulators but further simplifications will.

\subsection{Theory drift}

Early on Von Neumann and Wigner  noted that
some important evolutions are small homotopies
of the algebra.
Segal suggested the {\em simplicity principle}: 
Theories drift toward
(group) simplicity.
His reason is essentially Darwinian.
Our experiments  are disturbed
 by the many uncontrolled quantum variables of the
experimenter and the medium, so
our measurement of the structure tensor must err.
To survive a physical theory 
should be stable 
against small errors in the structure tensor.
Simple Lie algebras are stable Lie algebras.

This proposal rests on dubious implicit assumptions about
the domain of possibilities.
For instance, groups  that are stable in 
the domain  of Lie groups are 
unstable within the larger domain of quantum groups or non-associative products.
The group-stability criterion
might  produce some useful theories, but it 
might also exclude some.
Moreover there are a great many stable algebras that are not simple.

We infer the simplicity principle from the 
{\em general quantum principle}:
Any isolated system is a quantum system, 
with unlimited superposition.
Therefore the commutator algebra of its 
operational algebra is 
is the special orthogonal algebra
of
its io vector space (\S\ref{sec:QUANTIFICATION}),
which is simple.
A system whose algebra is not yet simple
is one we have
yet to resolve into its quantum constituents,
possibly because of strong binding or low resolution.

There are encouraging signs that when the
algebra becomes simple the theory becomes finite;
that
infinities today result from departures from algebra simplicity,
vestiges of classical physics that 
must be quantized.

The simplicity principle provides the kind of general understanding
of the development of physics that Darwin's theory of evolution
and Wegener's theory of continental  drift supply for biology and geology.
It does not predict the development
but suggests several possibilities 
for experiment to choose among.
It produces a phenomenological theory,
not a ``fundamental theory.''

One can illustrate such regularization-by-simplification 
with the same elementary example as before
 (\S\ref{sec:REGULARIZATION}).
The quantum linear harmonic oscillator 
has compound and singular Lie algebra 
and infinite-dimensional mode-vector space.
Its basic coordinate and momentum operators 
diverge on most of its mode vectors.
Segal stabilized this algebra by simplifying it, 
probably 
to $d\text{SO}(2,1)$,
though he did not state the signature explicitly.
This
has an irreducible representation $R(l)\text{SO}(3)$
of finite dimension $2l+1$ 
for any finite quantum number $2l=0, 1, 2, \dots$.
For all our finite experiments can tell us,
a sufficiently high-but-finite-dimensional  matrix representation
works at least as well as 
the infinite-dimensional singular limit.
Heisenberg's historic choice of the singular compound group 
$H(1)$ over the regular simple one $\0{SO}(2,1)$,
and Einstein's choice 
of the compound Poincar\'e group over a simple 
de Sitter group, natural as they were,
incorporated typical idol-formations of the kind that Bacon described.
The algebra simplifications described here
regularize the theories as well as stabilizing them
in some degree.
For another example where a well-chosen homotopy
replaces infinite-dimensional representations of a compound group 
by finite-dimensional ones of a simple group see \cite{KIM1987, KIM1990}.

In general, the irreducible representations of compound 
 algebras useful in
physics are nearly unique but contain serious infinities, 
while representations of
inearby simple algebras 
are numerous but finite.
It seems plausible that some of
these nearby finite-dimensional 
algebras suffice for present physics
at least as well as the present infinite-dimensional ones.

If the operational algebra of a system is not simple, 
it omits some  quantum degrees of freedom. 
As these are found and excited the theory will correspondingly 
simplify its Lie algebra.
Theory currently drifts  toward simplicity
because we are increasing the resolution
of our analysis of nature into quantum elements.
This is only one of the currents at work
in the drift of physical theory.

\subsection{The oldest game in town}

The deep changes in the structure of successful physical theories 
since 1900
--- special and general relativity, quantum theory, gauge theory --- 
introduced simplifications,
a mode of theory change that
lends itself to mathematical study.
The well-posed direct problem is
how
present theories contract to past  ones
\cite{ INONU1952}.
Segal worked on the ill-posed inverse problem
that concerns us here
\cite{SEGAL}.
How should present theories expand into 
future ones?  

Vilela Mendes \cite{VILELA1994} seems to have been the first
to apply the stability principle of Segal
to construct new quantum physical theories.
He noted that to simplify
most Lie algebras one must first introduce new variables
and  then invoke crystallization
to freeze them out in the vacuum.
He drew on the mathematical theory of stable 
(rigid) algebraic structures \cite{GERSTENHABER1964},
which in turn may have been influenced by Segal's proposal.

People have  since simplified the 
stationary theory of a  quantum harmonic oscillator 
 \cite{KUZMICH, Carlen2001, Atakishiyev2003, THOOFT,
 BAUGH2004,SHIRI2004} 
 and some its canonical dynamics
 \cite{BAUGH2004,SHIRI2005}.
Madore's ``fuzzy spheres''   include
 the Segal  simplification \cite{SEGAL} of the Heisenberg algebra 
 $d\0H(1)$
with 
one coordinate and one momentum as a special case
  \cite{MADORE1992, MADORE1999}.
  Golden has simplified a current algebra 
  in the manner of Vilela Mendes
  \cite{GOLDEN}.
  
The present  simplification program   
stems from  the stabilzation  program of Segal \cite{SEGAL}.  
(\S{sec:STABLE})
It extends the stabilization of space-time by Vilela Mendes
\cite{VILELA1994} to the higher level of dynamics.

\section{General quantization}
\label{sec:QUANTIZATION}

Singular theories  
are based on a Lie algebra  ${\3L}(0)$
that is compound, not semisimple,
and on a representation thereof --- call it $R(0) {\3L}(0)$ ---
that is not finite-dimensional.
For the canonical algebra $d\0H(1)$,  $qp-pq=ihbar$, $\dots$,
the 
representation  $R(0)$ is uniquely determined
by 
 unitarity, irreducibility,
and the value of one quantum constant $\hbar$
and is infinite-dimensional.

A {\em central invariant} is an algebraic
combination of Lie algebra elements that is central in every 
representation, and is therefore a c number in every irreducible representation.
For any Lie algebra ${\3L}$ 
and any representation $R$ of ${\3L}$,
the Casimir invariant $C_n$ of $R {\3L}$,
the coefficient of $z^n$
in the
characteristic polynomial 
\BE
C(z)=\text{det\,}(L-z\21)=\sum C_n z^n\quad \0{for }L\in R\3L
\EE
is a central invariant.
Planck's constant in the form $i\hbar$ is the value of the central invariant
$r=i\hbar$ for $d\0H(1)$.

It is convenient to introduce dimensional constants 
$\delta q_n$
to bring the generators $q_n$ of ${\3L}$
to a 
standard dimensionless form $L_n$ whose
spectrum has unit spacing. 
Then the
$C_n$ have integer eigenvalues $c_n$,
{\em quantum  numbers} that
define a representation algebra $R(\2c){\3L}$.

The algebra does not define its own physical meaning.
One must give its elements physical meaning.
If we double the value of $\hbar$ 
we change physical predictions
but do not change the algebra, up to isomorphism.
We may form a distinguished {\em physical basis} $\2B$
within the Lie algebra ${\3L}$, 
of physical variables $q_n$
defined by how we measure them
in standard units, not by their algebraic relations;
for example,  $q_1=$ position, $q_2=$ momentum, $\dots$.
{\em  Quantum constants} 
$\2h=\{\delta q_n \}$, 
including $\hbar$, then
define the representation $\2B(\2h)$ of the preferred basis
within the representation $R(\2c) {\3L}$.

To quantize  a singular theory in the present general sense we:
\begin{enumerate}
\item 
Warp its Lie algebras  to  simple ones
with as few new variables as possible.
\item Choose 
representations that
correspond with the singular theory in the experimental domain.
\end{enumerate}
The correspondence principle provides 
experimental meanings for some of the variables.

The ``space-time code'' 
\cite{FINKELSTEIN1969, FINKELSTEIN1996}
built quantum space-time  from the bottom up.
Vilela Mendes \cite{ VILELA1994, VILELA2005} and the present work build quantum space-time  from the top down.
The c space-time continuum
arises from STiME in
a singular limit of an organized mode of an underlying complex system,
the ether,
which determines no rest frame.
STiME splits into the usual fragments
--- space-time, the complex plane, and momentum-energy --- 
only relative to the ether.

The q space-time has  a basic kinematic symmetry between space-time 
and energy-momentum variables
like that  postulated   by Born and co-workers in their reciprocity theory 
\cite{BORN},
except that now it extends to $i$ as well,
whose
simplification $\Fi$
couples $p$ and $q$, $E$ and $t$.
This unification of time and energy 
violates common sense  even
more 
than the unification of time and space,
though this does not mean it is right.
The ether  condensation breaks this symmetry.

The present singular theory uses several singular algebras.
For example, classical mechanics
has both a commutative algebra of phase-space coordinates and
a Lie algebra of phase space coordinates
with the Poisson Bracket as product.
Classical space-time has a commutative algebra of
coordinates and a Lie algebra of vector fields with 
the Lie Bracket as product.
In such cases economy prefers a quantization that
deduces both singular algebras as singular limiting cases of one
more regular algebra,
as did canonical quantization.

Shall we follow the
orthogonal, unitary,  or symplectic line of simple algebras?
We work with huge dimensionality, so the exceptional algebras 
do not come in.
Experiment does not yet  clearly decide our choice.

I exclude the symplectic line because 
it lacks a well-behaved quantification theory.
The best-behaved quantification is that of the 
Clifford-Wilczek statistics of the real orthogonal D  line,
which enjoys the ``family property" \cite{WILCZEK1982}.
We take the D line here.
Baugh
takes the A line
\cite{BAUGH2004}.
It is too soon to say which agrees better with experiment.

\subsection{Terminology}
\label{sec:SIMPLIFYING}

Some terms:

A $\dag$ space $V$ is a vector space provided 
with an
involutory antilinear possibly indefinite anti-automorphism $\dag: V \to V^{\0D}$, 
the dual space.
The $\dag$ represents total time reversal
 \cite{FINKELSTEIN1996, SALLER2006}.
In a quantum theory, positive-signature unit vectors $|\psi\> \in V$,
$\psi\adj(\psi)=+1$,
 represent input modes;
positive-signature unit dual vectors  $\<\phi|\in V^{\0D}$ express output modes; 
the  transition amplitude  between them is
$A=\<\phi|\psi\>$.

We deal with both abstract and operational algebras or groups.
An operational algebra is an  algebra
 with an operational  interpretation.
The interpretation may be expressed by assigning names like
momentum or charge to elements, which define
how they are executed in the laboratory.
It suffices to do this for a basis. 
Similarly for groups.
Stretching some of the preferred basis elements does not change
the abstract algebra or the representation algebra but it
changes
the physical interpretation, and therefore the operational algebra.



A {\em stable}  $\dag $ Lie algebra is one 
whose Lie product $\4X: {\3L}\ox {\3L}\to {\3L}, a\ox b \mapsto [a,b]$  is  
isomorphic to all the Lie products $\4X'\in N(\4X)$,
a neighborhood of $\4X$ that are 
compatible with the same $\dag$.
(Segal's stability ignores $\dag$.)
\cite{SEGAL}. 
Because a non-singular $\dag$  is stable, we need not modify it.

Some concepts have  been formulated and named several times.
A stable algebra
 is variously and
synonymously said to be robust,  rigid,  
regular, or generic.
Conversely an unstable algebra is said to be fragile,  elastic,
singular, or special.
The operational algebras in a neighborhood differ slightly
in 
interpretation,
assuming that the distinguished basis is unchanged,
and so for physical purposes
Segal's  term ``stable''  is better than ``rigid.''

We can usually ignore 
the difference between the simple and semi-simple here. 
A direct sum is an incoherent mixture, and we see only one system at a time.
One well-chosen maximal measurement will reduce a semi-simple operational algebra to one of its simple ``superselected'' terms
for all subsequent measurements.

A homotopy $A_0\hookrightarrow A_1$
from one $\dag$ algebra $A_0$ to another $A_1$ (possibly Lie)
on the same $\dag$ space $A$,
each with its own product $\4X_0$ and $\4X_1$, 
 is
a continuous function $\4X: A\ox A \x I \to A$, where $I=[s_0,s_1]\subset \1R$
is an interval,
such that $\4X(a, a', s_0) = a\4X_0 a'$,
$\4X(a, a', s_1)=a\4X_1 a'$,
and each $\4X(a, a', s)=a\4X_s  a'$ is a $\dag$ algebra product for all $s\in I$\/.
Usually $s_0=0$.

A {\em simplification} is a  homotopy  $A\hookrightarrow A(s)$
 from a non-simple algebra $A=A(0)$ 
to a simpler algebra   $A(s)$ (say, with smaller ideals or nilradical), with homotopy parameter 
$s\in [0,s_1] \subset \1R$.
As an example Segal simplifies a canonical Lie algebra
$\0{Lie}(q, p, i)$ to a Lie algebra
of three generating angular-momentum-like variables 
$\0{Lie}(\8q, \8p, \8r)$,
replacing the central $i$ with the non-central $r$.
His homotopy transformation $L\x [0,s_1]\to L$
depends quadratically on
$s$.
The kind of homotopy that In\"on\"u and Wigner called a
contraction \cite{INONU1952}
is a special case that we
 call a  linear contraction.
Non-linear contractions are necessary 
 as well \cite{WEIMAR, SALLER2006}.
Linear contractions
sufficed to contract
special relativity to Galilean relativity 
and quantum theory to classical mechanics.
To recover classical mechanics from  canonical quantum mechanics
requires a
 quadratic contraction.
The regularizations of bosonic statistics and of space-time structure.
are inverses of  contractions in the 
more general non-linear sense.

\subsection{Quantum constants}
\label{sec:QCONSTANTS}

General quantization usually
introduces 
new regulation operators or {\em regulators} $q_n$
and several kinds of physical constant:
\begin{enumerate}
\item   Signatures defining the simplified Lie product 
 $\8{\4X}$. 
 \item Regulation constants or {\em regulants} $\6q_n$,    
expectation values of regulants $q_n$ in the ambient ether, setting 
physical scales.
 \item {\em Quantum numbers} $\2c$ defining a representation
 $R(\2c): {\3L}\to A(\2c)$ of the simplified Lie algebra.
 \item {\em Quantum constants} $\delta q_n$ defining the spectral intervals of physical operators $q_n\in A(\2c)$.
 \end{enumerate}

The regulants $\6q_n$ are typically both spectral  maxima and ambient values  of regulators,
\BE
\max |q_n|=<\0{vac}|q_n|\0{vac}>:=\6q_n.
\EE
 We  simplify  the
canonical relation $pq-qp=-i\hbar$
to the  cyclic form
\BE
\8p\8q-\8q\8p=\frac{\delta p\, \delta q}{\delta r} \8r, 
\quad \text{\& cyc}\/
\EE
on dimensional grounds.
The operator that
freezes to $-i\hbar$ in the singular theory is 
clearly 
\BE
\Fi \hbar=\frac{\delta p\, \delta q}{\delta r} \8r:=N \delta p\, \delta q.
\EE
where the integer $N$ is the maximum eigenvalue  of $|\8r|$ 
as a multiple of its quantum $\delta r$.

Canonical quantization and special-relativization
introduced scale or quantum constants but no regulators.
Subsequent simplifications have both 
\cite{SEGAL,VILELA1994, BAUGH2004, SHIRI2004, SHIRI2005}.

Theories are c or q as their dynamical variables 
all commute or not. 
Then
q theories divide into q/c and q/q
as their time is commutative or not
\cite{FINKELSTEIN1996}.
We formulate a q/q physics here

The two main ways to formulate a c dynamical theory,
Hamiltonian and Lagrangian,
have q/ and q/q  correspondents.
When  the Hamiltonian theory
is singular,
the Lagrangian theory is even more so,
because the system  has many more history modes than  
single-time modes.
I simplify the Lagrangian theory.

In a q/q theory, the algebras of all levels within the theory
are non-commutative.
To regularize such hierarchic theories we must regularize all their
constituent algebras
and the algebraic relations between levels.
For this we use an algebraic concept of quantification (\S\ref{sec:QUANTIFICATION})
or statistics.

\section{Quantification}
\label{sec:QUANTIFICATION}

The passage from a one-system theory to a many-system theory
is 
a general process aptly named quantification
by the Scottish logician
William Hamilton (1788- 1856).
It is not a quantization and is much older.

The operations of system creation and annihilation
can be represented as withdrawals and deposits from a reservoir
of like systems.
The one-quantum theory keeps the reservoir off-stage and
represents
the io actions not by operators but only by vectors,
halves of operators.
Quantification brings the reservoir on-stage and represents these io actions by
operators, not mode vectors.
The notion that experiments on a single quantum
can tell us the operational algebra of a many-quantum system
is a relic of c physics,
where bodies are made of atoms and
the body state space is the Cartesian product 
of many atom state spaces,
but it might still be right,
and has worked amazingly in the q theory,
with necessary changes.

The earliest quantum physicists naively took for granted
that 
the many-quantum mode space 
is the tensor algebra over $V$,
$\0{Tensor }V$\/.
This is Maxwell-Boltzmann statistics,
the quantification
 for 
fictitious quanta that we can call maxwellons.
Then Bose realized that the io operators generating the quantified
algebra must obey significant commutation relations.
For example the bosonic and fermionic quantifications
 $\Sigma_{\sigma}$ 
are based
on the algebraic relation
\BE
\label{eq:3QUANTIFICATIONS}
b\adj a = \sigma a b\adj +\<b|a\>
\EE
for $a,b\in V$.
We confine ourselves here to the algebras with
$\sigma = +$ (bosonic), 
$\sigma = -$ (fermionic), $\sigma = 0$ (maxwellonic)
and their simplifications and iterations.

The Lie algebra (\ref{eq:3QUANTIFICATIONS}) for $\sigma=+$ is a singular canonical algebra $d\0H(D)$.
We simplify it to a regular
orthogonal algebra $d\0{SO}(2D+2)\hookrightarrow d\0H(2D)$.

Standard quantum theory forbids the superposition of bras and kets
and bars total time reversal from the operational algebra.
This rule seems to be a c vestige in the present quantum theory.
We break it and form the direct sum of the two io spaces of bras $V$ and kets
$V^{\0D}$ \cite{SALLER2006},
calling this the io (vector) space 
\BE
W=V\oplus V^{\0D}.
\EE

By
a quantification $\Sigma$ 
I mean a construction  that
sends  each one-quantum io space $W_1$
 to a many-quantum io space
$W=\Sigma W_1$
depending on
\begin{itemize} 
\item 
the structure $\4X$ of a 
$\dag$ Lie algebra $\3L(W_1, \4X)$ on $W_1$,
\item 
 an irreducible 
representation  $R:\3L(W_1, \4X)\to A$,
and 
\item a vacuum projector $P(\0{vac})\in A$
\end{itemize}
The quantified  io vector space is then defined as
\BE
\label{eq:LQUANTIFICATION}
W=\Sigma W_1= R\3L(W_1, \4X) \; P(\0{vac})\/.
\EE
 If we diagonalize $P(\0{vac})$,
all the operators
$\Sigma W_1 \; P(\0{vac})$
are matrices with only one non-zero column
which represents a ket.

To define a quantification we must give not only
the Lie algebra
but must also give values
$\2c = \{c_n\}$ 
for its central (Casimir) invariants $\2C$ to
define an irreducible
representation  $R(\2c): {\3L}\to A$\/,
the endomorphism algebra of the
$\dag$ space $W$. 

We regard   $\Sigma_{\4X}$ 
for any Lie algebra $\3L=(W,\4X)$
as a generalized statistics 
or quantification 
for quanta that we can call ``${\3L}$-ons.''
Bosons and fermions result from Bose and Fermi 
(graded Lie) algebras respectively,
which are canonical and Clifford algebras
respectively.
In the fermion case the input and output vectors in $W$
are required to be null vectors in the Clifford algebra.
This is a singular condition
that has already been regularized for other reasons
\cite{FINKELSTEIN1982, WILCZEK1982, FINKELSTEIN1996}.
We   regularize the bosonic statistics  in 
\S\ref{sec:REGULARFIELDALGEBRA}.

Since the  classical continuum is singular,
we 
regard all our Lie algebras as ultimately statistical.

Dynamics has a hierarchy of at least five algebras 
(\ref{tab:ALGEBRAS}).
In formal logic such hierarchies  are handled with quantifiers.
In q/c physics the lower level c quantification is handled informally and intuitively,
and the higher q level quantification
is constructed from the lower algebraically as in \S\ref{sec:QUANTIFICATION}.
In q/q physics we must handle all quantifications algebraically.
  
Quantification
 deceptively  resembles quantization
in more than spelling.
 Both adduce commutation relations,
 and they may even end up with the same algebra.
 Nevertheless they are conceptual opposites and
 if they come to the same place, they arrive there
 from opposite sides.
 Quantification sets out from a one-quantum theory.
 Quantization set out from a classical theory,
 which is a many-quantum system seen 
 under low resolution and
 with many 
 degrees of freedom frozen out.
For extremely linear systems like Maxwell's,
the two starting points may have similar-looking
variables but the operational meanings of those
variables are as different as c and q.

\section{Finiteness and stability}
\label{sec:FINITE}

Simple Lie algebras seem to result in finite (= convergent)  theories.
We begin to explore this delicate question here.
Compact simple Lie algebras  have complete sets of finite-dimensional representations 
supporting
finite-dimensional quantum theories  with no room for  infinities.
The simple algebras with indefinite metric required for physics
have problematic 
infinite-dimensional irreducible unitary representations
besides the good finite-dimensional ones.
We hypothesize that we can  approximate
the  older singular compound theory
without these infinite-dimensional representations;
this has been the case for the Lorentz group, for example,
and it is consistent with
analytic continuation from the compact case of positive definite $\dag$.

Regularity also divides mechanical theories with singular  Hessian determinants 
from those with regular Hessians.
Indeed, all singularities that depend on some variable 
determinant miraculously vanishing 
are  non-robust, non-generic, unstable by that fact,
and are eliminated by general quantization.

\subsection{Regularization  by simplification}

The Lie algebraic products, or structure tensors, $\4X: V\ox V\to V$ admitted by a given vector space $V$,
 form a quadratic submanifold  $\{\4X\}$ in the linear space of tensors over $V$,
defined by the Lie identities 
\BE
\4X(a\ox b+b\ox a)=0, \quad\4X^2 (a\ox b\ox c +c\ox a\ox b+b\ox c\ox a)=0.
\EE
The quotient of this manifold by the equivalence classes modulo Lie-algebra isomorphism is the moduli space of Lie algebras on $V$ \cite{FIALOWSKI}.

Any singular  Lie algebra  lies on the  lower-dimensional boundary in $\{\4X\}$
of a finite number of these classes.
For example, 
the 6-dimensional Galilean algebra of rotations and boosts sits between the $\0S\0O(4)$ algebras and the $\0S\0O(3,1)$ algebras.
 To simplify such a singular algebra we merely
 move its structure tensor off this boundary to 
a nearby simple algebra
\cite{SEGAL, GERSTENHABER1964, VILELA1994, SALLER2006}.
The simple group approximates the compound one
only near their common point of tangency,
as a sphere approximates a tangent plane,
in a correspondence domain
whose size is set by 
a physical constant or constants new to the
singular theory,
and which must include the experiments
that have been satisfactorily described by the singular theory.
A simplified theory $\8\Theta$,
by
fitting its regulants into 
the error bars of the unsimplified theory
$\Theta$\/,
 inherits the operational semantics
and past experimental validations of  
$\Theta$,
while still making radically new theoretical
predictions about future experiments,

\subsection{Regulators} 
\label{sec:REGULATORS}

If we introduce regulators we also need to explain
how the unregulated singular theory
could work as well as it does without them.
Call the subspace of the regular mode-vector space where the regular theory
 agrees with the singular theory within experimental error,
 the {\em correspondence domain}\/.
We hypothesize that
self-organization 
produces a gap that freezes out the regulators 
 in the correspondence domain,
 where the singular theory gives some good results.
 Self-organization is also responsible for the emergence 
 of classical mechanics from quantum mechanics.
 
General quantization
exposes a larger symmetry algebra,
supposed to have been hidden in the past
by self-organization,
and able to manifest itself in the future
 under  extreme conditions
like ether melt-down.
Carried  far enough,  general quantization
  converts a singular theory with a compound 
(= non-semisimple) algebra
into a regular theory with a simple algebra
\cite{SEGAL}.
This requires no change in the stable elements of a theory,
only in the unstable elements, such as 
the classical theory of space-time.

Suppose that the simple Lie algebra is an orthogonal one 
$d\0{SO}(N)$
(rather than unitary or symplectic).
 Then we can choose each simplified generating variable $q$  
 to be a multiple of an appropriate dimensionless component  
 $L^{\alpha}{}_{\beta}$ 
 of an angular momentum in $N$ dimensions, by a 
 dimensional constant $\delta q$:
 \BE
 \label{eq:QCONSTANT}
\8q = \delta q  L^{\alpha}{}_{\beta}.
\EE
  We adjust the spectral spacing of $L^{\alpha}{}_{\beta}$ to  1.
Then the quantum of $q$ is $\delta q$.
  To diagonalize an antisymmetric generator $L^{\alpha}{}_{\beta}$ requires
adjoining a central $i$ for the purpose.
Then the generators are all quantized with 
uniformly spaced,  bounded, discrete spectra.
The  maximum of 
the absolute values of the eigenvalues of $\8q_n$ 
we designate by $\0{Max}\8q_n$.

 These $\delta$'s generalize  the quantum of action, 
 $\delta A=\delta (E/\omega)= \hbar$,
so we call them quanta of their variables.
For example, simplification introduces quanta 
$\delta x$ of position, $\delta t$ of time, 
$\delta p$ of momentum, and 
$\delta E$ of energy, as well
 as the familiar quanta of charge and angular momentum.

The main singular algebra of q/c physics,
the Heisenberg algebra
$d\0H(M)$ (for $M$ spatial dimensions),
whose radical includes $i\hbar$\/,
has already been simplified for $M=1$
 \cite{SEGAL, VILELA1994, KUZMICH, Carlen2001,Atakishiyev2003, 
 BAUGH2004,SHIRI2004,THOOFT}
and for $M>1$, both  unitarily \cite{BAUGH2004} and orthogonally \cite{SHIRI2004, VILELA2005}.

\section{A regular relativistic dynamics}

Can the  infinite-dimensional representations
of the non-compact groups like the Poincar\'e group
that are used in quantum physics today
indeed be 
 approximated by  finite-dimensional representations of an
 approximating orthogonal group?
 In the non-compact cases the orthogonal groups have infinite-dimensional 
 irreducible unitary representations as well as finite-dimensional orthogonal ones.
The danger is that an infinite-dimensional representation is required for this approximation, with its native divergences.
 
A typical example: Consider
 a scalar quantum of mass $m$ in a space-time of 
$3+1$ dimensions.
One  can approximate its singular Poincar\'e Lie algebra 
$d\text{ISO}(3,1)$ 
with a regular de Sitter Lie algebra 
$d\text{SO}(5, 1)\to d\text{ISO(3,1)}$.  
A scalar massive quantum in Minkowski space-time provides
an infinite-dimensional  unitary representation  
$R \,d\text{ISO}(3, 1)$ 
 in use today.
Can one approximate this useful infinite-dimensional representation of the singular algebra
 by  a finite-dimensional representation  of the regular algebra?

At least five major  Lie algebras arise in such models:
 \begin{center}
 \label{tab:ALGEBRAS}
  \begin{tabular}{@{}llll @{}}
 Level &Space & &Lie algebra\\
\hline
1 &  Space-time tangent space&    $d\3X=\{dx\}$          &   $d\text{SO}(3,1)$  \\ 
  2&Space-time&  $\3X=\{x\}$                                            &  ${\3L}_{\3X}$  \\ 
  3& Field-value tangent space& $d\3F=\{df\}$& ${\3L}_{d\3F} $\\
  4& Field-value space  &$\Phi=\{\phi\}$  & ${\3L}_{\Phi}$ \\ 
 5&  Field history space &$ \3F=\{f\}$                    & ${\3L}_{\3F}$  \\ 
\end{tabular}
\end{center}
The initial hierarchic structure is a lambda
we assume,
with space-time and field variable on the same level,
and the final structure is simpler:
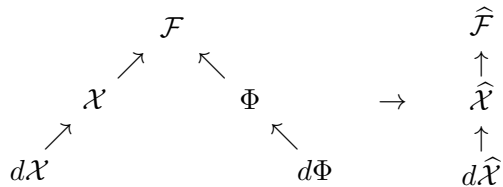
\begin{center}
 \label{tab:LEVELS}
  \begin{tabular}{c}
           \kern20pt$\3F$\kern20pt\\ 
$\nearrow$\kern20pt$\nwarrow$\\
$\3X$\kern50pt$\Phi$\\ 
$\nearrow$\kern75pt$\nwarrow$\\
$d\3X$ \kern90pt$d\Phi$
\end{tabular}$\quad \to\quad  $
  \begin{tabular}{c}
$\8{\3F}$\\ 
$\uparrow$\\
$\8{\3X}$\\ 
$\uparrow$\\
$d\8{\3X}$ 
\end{tabular}
\end{center}

The Lorentz algebra ${\3L}_{d\3X}$ is regular 
and for the scalar field ${\3L}_{\Phi}$ and ${\3L}_{d\3F}$ are the one-dimensional Lie algebra.
We regularize  the remaining algebras here.

\subsection{Regular space-time}
\label{sec:RSPACETIME}

We simplify space-time first,
then the scalar field on that space-time.
This is mainly an illustrative example
chosen as training for the most interesting
singularity,
that of gauge theory and
 gravity,
which
suggests a quantum space-time
that we take more seriously.

The usual space-time coordinates $x^{\mu}$ 
 generate a compound commutative 
four-dimensional Lie algebra.
There is no  4-dimensional simple Lie algebra.
To make simplification possible
without losing Lorentz invariance
 we 
adjoin  the four differential  operators $\partial_{\mu}$ and 1
 as regulators,
resulting in the compound Lie
algebra $\0H(4)=\text{Lie}(x^{\mu}, \partial_{\mu}, 1)$\/
with standard commutation relations understood.
This may also be the most economical way.

 Now the irreducible unitary representation is essentially unique:
 The generators $x^{\mu}, \partial_{\mu}, 1$ 
  act in the standard way on $\0L^2(\3M^4)$.
This is also isomorphic to the diachronic pre-dynamical 
operational Lie algebra of a single scalar quantum particle in space-time.
Statements about position in the abstract have been imbedded
in statements about a quantum
particle of unspecified dynamics,
which we call ``the probe.''
Inevitably this brings in statements about the momentum
of the probe  as well.
This is but a partial regularization of space-time,
neither regular nor simple.

$\text{Lie}(x^{\mu}, \partial_{\mu}, 1)$ is also the Lie algebra 
$\Sigma_+ V(3,1)$
of a certain bosonic aggregate.
The mode-space $V(3,1)$ of the individual boson 
is isomorphic to the tangent space $d\3M^4$
to four-dimensional Minkowski space $\3M^4$ at the origin
but is not that space, being
interpreted in a way that is non-standard for differential geometry.
Its vectors  are mode-vectors of a hypothetical quantum;
the ``minkowskion,''  let us call it.
The classical space-time is now presented, 
ready for regularization,
 as a 
bosonic aggregate of minkowskions
which has been
reduced to a classical system 
by freezing out the momentum-energy variables,
and centralizing (``superselecting'') 
 the coordinates $x^{\mu}$,
effectively restricting frames to the classical
 space-time coordinate basis $|x^{\mu}\>$.
 No quantum of space-time has entered yet,
 but  quantum variables have.
To take  quantum space-time seriously
one must eventually find a physical mechanism that freezes 
half the variables
by self-organization
 (\ref{sec:REGULARFIELDALGEBRA}).

Now we simplify fully.
This calls for more regulators.
We follow the D line and 
adjoin 6 Lorentz generators
$L^{\mu}{}_{\nu}=-L_{\nu}{}^{\mu}$
to the present generators $x^{\mu}, \partial_{\mu}$,
assuming  a fixed background Minkowski metric $\dag$
that interchanges vectors and dual vectors,
raising and lowering indices.
This expands the 9-dimensional 
canonical Lie algebra $d\0H(4)$ to a still singular
15-dimensional Lie algebra 
$\text{Lie}(x^{\mu},\partial_{\mu},
L^{\nu}{}_{\mu}, 1)$ with the commutator $AB-BA$ as Lie product $[A, B]$ and
with  standard commutation relations (\ref{eq:LIE15}) 
for these operators.
This algebra can be simplified to 
a 15-dimensional orthogonal algebra $d\0{SO}(6)$
of signature to be determined.

This simple space-time is more quantum than the Snyder space-time,
which is not simple.

Notation:
We label simplifications by a collective argument 
$\2q=
(\2h,\2c)$ with $\2h=\{\delta q_i\}=\2{\delta q}$
 consisting of quantum constants like $\hbar$ and $1/c$, and 
 with 
  $\2c=\{c_n\}$ consisting of quantum numbers   defining
 values 
 of all central invariants 
 (see \S\ref{sec:QUANTIZATION}).
The passage to a singular limit we write as $\2q\to \2q_0$.
  We absorb factors of $i$ to make the variables  
 $q_i$ anti-Hermitian for convenience.
 We may omit the circumflex that indicates  simplification 
when it is  redundant.
The old indices
$\mu, \nu = 0, 1, 2, 3$   label space-time 
or momentum-energy axes in the singular theory. 
Special constant index values 
$\0X,\0Y$  label real and imaginary units
in the complex plane of the singular limit. 
They
 distinguish space-time variables $L_{\mu\0X}$ 
 from momentum-energy variables 
 $L_{\mu \0Y}$
  in the regular theory.
Extended indices $\alpha, \beta=0, 1,2, 3, \0X, \0Y$ 
 label axes in the orthogonal space that supports 
the regular group $\text{SO}(5, 1)$.
We may set $\hbar=c=1$ 
since we hold them constant as $\2q\to \2q_0$.
$\3X(\2q_0)$ is the singular quantum space 
and  the associative algebra
defined by the usual infinite-dimensional unitary representation
of its Lie algebra
\BE
{\3L}_{\3X}(\2q_0):= \text{Lie}(x^{\mu},p_{\mu},
L^{\nu}{}_{\mu}, i)
\EE
 on the function space $\0L^2(x^{\mu})$.
${\3L}_{\3X}(\2q_0)$  has the familiar singular structure 
\BE
\label{eq:LIE15}
\begin{array}{llrllrll}      
 [x^{\nu}, x^{\mu}]=0,  & [x^{\nu}, p_{\mu}] = i\delta^{\nu}_{\mu}, &   
 [x^{\nu}, L_{\mu\lambda}]&=&\delta^{\nu}_{\mu} x_{\lambda}-
 \delta^{\nu}_{\lambda} x_{\mu}, & 
 [x^{\mu}, i ]&=& 0,\\
& [p^{\nu}, p^{\mu}]=0, &[ p^{\nu}, L_{\mu\lambda} ]&=&
\delta^{\nu}_{\mu} p_{\lambda}-
 \delta^{\nu}_{\lambda} p_{\mu}, &[p^{\mu}, i ]&=& 0, \\
&&[L^{\nu\mu}, L_{\lambda\kappa}]&=&
\delta^{[\nu}_{[\lambda}L^{\mu]}{}_{\kappa]}, &[L_{\nu\mu}, i]&=&0\\
\end{array}
\EE

The singular Lie algebra ${\3L}_{\3X}(\2q_0)$ simplifies to a regular Lie algebra $\8{\3L}_{\3X}(\2q)\sim d\text{SO}(5,1)$ or 
$d\text{SO}(3,3)$ as follows
\cite{VILELA1994}.

First the idol $i$ melts down to the Lie element
$\Fi:=\8r/\0{Max}\,\8r$\/.
We return to the singular theory by freezing  $\Fi$
at its maximum eigenvalue.

We choose the Minkowskian signature
to postpone the problems of multiple timelike axes.
Then  the dimensionless infinitesimal orthogonal transformations 
$L_{\beta\alpha}\in d\text{SO}(5, 1)$ rescale
to  simplified versions  of the generators 
of ${\3L}_{\3X}(\2q_0)$ in ${\3L}_{\3X}(\2q)$ .
The  15 variables $L_{\beta\alpha}$ require
 four quantum constants $\2h=(\delta x, \delta p, \delta L, \delta r)$,
 but $\delta L=\hbar=1$ for Lorentz invariance:
\BE
\8L_{\nu\mu}= L_{\nu\mu}, 
\quad
 \8x_{\mu}  = \delta x\, L_{\mu \0X}, 
\quad 
\8p_{\mu}=\delta p \,L_{\mu {\0Y}},
\quad
\8r = \delta r  \,L_{\0X{\0Y}}.
\EE
The maximum eigenvalue of $-(L_{\alpha\beta})^2$ is the same for any spatial
$(\alpha\beta)$   plane,
a new quantum number we write
as $l_{\3X}^2$. 
Evidently in the singular limit we must have
\BE
\delta x \,\delta p=l_{\3X} \,\delta r \,\hbar
\EE
and we might as well impose this in general.

 This simplification
 converts the compound Lie algebra ${\3L}_{\3X}(\2q_0)$ 
 to a simple  Lie algebra ${\3L}_{\3X}(\2q)$ 
with generators $L_{\alpha\beta}$. 
The canonical commutation relations survive in the simplified form 
\BE
[x^{\mu}(h), p_{\nu}]=\delta^{\mu}_{\nu}
\frac{\delta x \delta p}{\delta r}
r\/.
\EE

We construct a quantum space STiME $= \8{\3X}=\3X(\2q)$
from its Lie algebra ${\3L}_{\3X}$
by
specifying an irreducible  matrix
represention $R(\2h) {\3L}_{\3X}$,
whose algebra is then the operational algebra of $\8{\3X}$.
The singular space-time algebra is 
an infinite-dimensional irreducible unitary representation
$R(\2h_0){\3L}_{\3X}(\2c_0)$
supported by the function space $L^2(\3X(\2c_0))$.
To fix on one regularized STiME
we must fix on quantum constants and quantum numbers $\2q$ 
of the Lie algebra ${\3L}_{\3X}$,
defining a preferred basis $\2B$ in
one irreducible representation
$R(\2h) {\3L}_{\3X}(\2c)$.

And the Lie algebra ${\3L}_{\3X}$ is
specified in turn by signatures.

The quadratic mode-vector space 
supporting  the defining representation of
$ {\3L}_{\3X}$ is
a 6-dimensional space $V_{\3X}$\/.
We form a high-dimensional representing vector space 
$R(\2h) V_{\3X}$ 
with collective quantum constant ${\2h}$\/,
to support the physical representation $R(\2h) {\3L}_{\3X}$. 
The singular space is spanned by polynomials in the coordinates,
and limits thereof.

An irreducible representation  $R({\2c})\0{SO}(5,1)$
 is defined by eigenvalues $c_n$ of the
Casimir invariants $C_n$, 
the coefficient of $z^n$
in the invariant 
characteristic polynomial 
\BE
C(z)=\text{det\,}(L-z\21)=\sum C_n z^n
\EE
 for $L\in dR({\2h})\0{SO}(5,1)$.
$C_n$   vanishes for odd $n$ because $L\sim -L$, leaving 
$C_2, C_4, C_6$.
As usual  $iL_{\0X\0Y}$ has  eigenvalue spectrum of the form
$-{{l}}, -{{l}}+1, dots, {{l}}-1, {{l}}$.
The extreme value $l$  is a regulant and $2{l}$ is an integer.
In the singular limit $l\to\infty$.

Let $\4L=(L^i_j)$ be the matrix whose elements
are the infinitesimal generators of $dR({\2c}) \0S\0o(5,1)$;
a matrix of matrices.
Then for each $n\in \1N$\/, $\0{Tr\,} \4L^n$ is another convenient 
invariant, whose value in the chosen representation we designate
by $\Lambda^{(n)}$. In particular,
\BE
\Lambda^{(2)}=-(L_{\0X{\0Y}})^2 -L^{\0X\mu}L_{\0X\mu}-L^{{\0Y}\mu}L_{{\0Y}\mu} +L^{\nu}{}_{\mu}L^{\mu}{}_{\nu}
\EE
The cross-terms $-L^{\0X\mu}L_{\0X\mu}-L^{{\0Y}\mu}L_{{\0Y}\mu} $ have vanishing expectation for any eigenvector of $L_{\0X{\0Y}}$
 by the generalized uncertainty inequality.
Then
\BE
\Lambda^{(2)}= l^2 -(\delta x)^{-2}\<\8x^{\mu}\8x_{\mu}\>
-(\delta p)^{-2}\<\8p^{\mu} \8p_{\mu} \>+\< L^{\nu}{}_{\mu}L^{\mu}{}_{\nu}\>
\approx  l^2
\EE
holds for the vacuum, as an eigenvector of extreme $L_{\0X{\0Y}}$.
In the correspondence domain one may drop the circumflexes.

This is a simplified Klein-Gordon equation with a  ``mass'' term
that depends on the STiME cooordinates and angular momentum.
Wigner taught us that the scalar fields supporting  irreducible representations of the Poincar\'e
group obey Klein-Gordon wave equations.
Naturally  a simplified group leads to a simplified wave equation.

Similarly 
\BE
\forall n\in \1N\;\Bar\quad {c}^{(2n)}-{l}^{2n}\approx 0 = {c}^{(2n+1)}
\EE
are polynomial conditions  on $\8x^{\mu}$, $\8p_{\mu}$, and $ L^{\mu}{}_{\nu}$
with coefficients depending on $\2h$ and ${l}$. 

Raising the dimension of the group has increased the number 
of invariants and wave equations.

The algebra $\8A=\8A (\8{\3X})$ 
of coordinate variables of the regular
STiME quantum  space 
$
\8{\3X}:=\3X(\2h)
$
is
the operator algebra  of the 
vector space $R(\2c) V_{\3X}$ that we have just constructed:
\BE
\8A:=\text{Endo }R(\2c) V_{\3X}\/.
\EE

Each factor in $R(\2c) V_{\3X}$ contributes angular momentum 
$\pm 1$ or 0  to each generator $L_{\alpha\beta}$
of ${\3L}_{\3X}(h)$, so the eigenvalue of $i R(\2c)L_{\alpha\beta}$
varies from $-l$ to $l$ in steps of 1.
Now the space-time coordinates and the energy-momenta are unified 
under the Lie group generated by $R(\2c){\3L}_{\3X}$ .
Each has
a discrete bounded spectrum with $2L+1$
 values $x= i \delta x \, m$, $p=i \delta p \, m$, for
$m\in \1Z, |m|\in l+1\/.$
Both operators are elements of the  STiME operator algebra
$A \3X_h:=\text{Endo }R(\2c){\3L}_{\3X}$, which replaces $\0L^2(\3X_0)$.

The regular quantum point of STiME  can be represented as a series
of $l$  more elementary processes or chronons, 
all identical,
a bosonic ensemble constrained to a fixed number $l$ of elements.
The chronon is a minkowskion in this model. 

Next
we  set up a singular scalar q/c field theory 
on the singular quantum space-time
so that we can regularize it in \S\ref{sec:REGULARFIELDALGEBRA}.

\subsection{Singular field Lie algebra}
\label{sec:SINGULARFIELDALGEBRA}

We label  the singular q/c limit with a suffix $(\2q_0)$ 
 and 
generic q/q case with $(\2q)$, dropping the circumflex, where $\2q$ 
is a collection of quantum constants  and quantum numbers
to be specified.

In the c scalar theory a history $f$ of the field is a pair 
$(f(\cdot), p_f(\cdot)$
of a field  function $f: {\3X}\to \1R$ on  space-time, and a 
contragredient momentum function $p_f:  {\3X}\to \1R$.
The space of such c histories is, aside from continuity requirements,
\BE
\3F=\1R^{\3X}=:\0D\3X,
\EE
a kind of linear dual of $\3X$.
The c functional Lie algebra is commutative:
\BEA
{f}(x){f}(x')-{f}(x'){f}(x)&=&0,\cr
{p_f}(x){p_f}(x')-{p_f}(x'){p_f}(x)&=&0,\cr
{p_f}(x){f}(x')-{f}(x'){p_f}(x)&=&0,\cr
{f}\adj+{f}&=&0,\cr
{p_f}\adj+{p_f}&=&0.
\EEA

The bosonic aggregate, or the quantum field,
has the formal functional Lie algebra 
 $L_{\3F}(\2q_0)$ 
generated by the operators on $\0D\3X$ of
multiplication by $f(\cdot)$
variational differentiation
$p_f=\delta_f(x):=\delta/\delta f(x)$, 
and the central $i$,
subject now to the canonical relations
of a $d\0H(\infty)$,
\BEA
{f}(x){f}(x')-{f}(x'){f}(x)&=&0\cr
{p_f}(x){p_f}(x')-{p_f}(x'){p_f}(x)&=&0\cr
{p_f}(x){f}(x')-{f}(x'){p_f}(x)+i\hbar\delta(x-x')&=&0\cr
{f}\adj+{f}&=&0\cr{p_f}\adj+{p_f}&=&0.
\EEA
for all $x, x'\in \3X$.
We  make both ${f}$ and ${p_f}$  anti-Hermitian
with incorporated factors of $i$ where necessary,
for the sake of the development to come.

Because of the two-story construction
the operators $x^{\mu}$ and $\partial_{\mu}$
in the space-time Lie algebra $d{\3L}_{\3X}$
can also act on the field functional Lie algebra $d{\3L}_{\3F}$,
with obvious commutation relations.

 The element $i\hbar$  is a 
 complete set of central invariants of this 
 functional Lie algebra.
The canonically quantized scalar field is a bosonic aggregate of individuals 
 whose mode-vector space is $\0L^2(\3M)$.

 This is the theory we simplify  next.
 
\subsection{Regular field Lie algebra}
\label{sec:REGULARFIELDALGEBRA}
The q/c scalar field is a bosonic aggregate.
The Lie algebra of bosonic statistics is  unstable, compound.
We 
simplify it now to a simple, stable, and finite near-bosonic statistics.

We use the Lie-algebraic procedure $\Sigma_{\3L}$ of (\ref{eq:LQUANTIFICATION}).
A fixed io mode-vector space $V$
 for an individual quantum $\0I$ is given,
 and we give a Lie algebra  ${\3L}$ on $V$, with 
 a structure tensor $\4X$
 close to the bosonic.
It is convenient to give $\4X$ on an $\iota$-labeled replica of $V$ 
in case there
 are other Lie algebra structures
 already defined on $V$,
 as in multiple quantification.

Then the mode-vector space $\Sigma V$ 
of the quantified system is
determined by the Lie algebra ${\3L}$ and quantum
constants $\2c$.
A vacuum projection $P(\0{vac})\in \Sigma V$
then determines the
vector space $\Sigma V \; P(\0{vac})$ as a 
mode-vector space for the quantified system.

In the case of
the singular q/c boson quantification 
${\3L}$ is the functional Lie algebra $\0{Canon}_+\iota\adj V$,
defined by 
the bosonic commutation relations
 on the union $V=V_{\0I}\cup V_{\0O}$
 of the input and output mode-vectors of the system.
 $V$ is a partial vector space; addition works within
 each term but not between them.
 
Then the bosonic operational 
algebra of the aggregate  of individuals I
is the target algebra $A(\2c)$
of the irreducible unitary representation $R(\2c)$
of ${\3L}$ with the central invariant $c=i\hbar$ specified. 

The usual creator and annihilator  of the many-quantum (or quantified) theory
associated with the mode-vectors $v$ and $v\adj$ of the one-quantum theory
are left multiplication 
 $\iota \adj v$ and  differentiation
 $(\iota \adj v)\adj=v\adj\iota$
 with respect to  $\iota \adj v$\/.

 If the $v_n\in V$ form a basis of input vectors with dual 
 output vectors,
 the corresponding creation/annihilation operators 
 $a_n:=\iota\adj v_n$,
 $c^n:=v^n\adj\iota$ obey
 \BEA
 \label{eq:NBOSECR}
 c^n a_m-a_mc^n-i\hbar \delta^n_m&=&0,\cr
 c^n c^m-c^mc^n&=&0,\cr
 a_n a_m - a_m a_n&=&0.
 \EEA
 This algebra is doubly infinite-dimensional: once 
 because bosonic quantification turns each dimension
 on the one-quantum space into an infinity of dimensions
 in the many-quantum space,
 and once because the one-quantum space has an infinity of dimensions, because space-time is infinite and continuous.
 The  space-time infinity is again bosonic, 
 arising from the fact that space-time is
 a bosonic aggregate of minkowskions,
 
A boson Lie algebra on a $2N$ dimensional io vector space is a contraction
of an SO Lie algebra on $2N+2$ dimensions.
To  simplify the relations (\ref{eq:NBOSECR})
 we first transform bosonic variables $a_n, c^n$ to  canonical anti-Hermitian variables
$q_n=-q_n\adj$, $p_n=-p_n \adj$ 
($n\in \1N$) using the  imaginary unit $i$:
\BE
a_n= \frac  {q_n/\delta q +ip_n/\delta p} {\sqrt 2}, 
\quad 
c_n= \frac  {q_n/\delta q -ip_n/\delta p} {\sqrt 2}, 
\EE
including quantum constants $\delta q, \delta p, \delta r$ for dimensional reasons.
Then
we introduce two extra real dimensions with indices  $\0X', {\0Y'}$
forming  a real vector space $V\oplus \22$ with 
vector indices $\alpha, \beta= 0, \dots, N-1, \0, {\0Y}$\/.
A symmetric metric $\dag: (V\oplus \22)\to (V\oplus \22)^{\0D}$  
defines  an orthogonal Lie algebra $d\0S\0O(V\oplus \22)$ generated 
by $(N+2)\x(N+2)$ matrices $L_{\beta\alpha}$, anti-Hermitian with respect to $\dag$\/.
We  represent the simplified simple-bosonic creators and annihilators   in the simple Lie algebra
$d\0S\0O(V\oplus \22)$:
\BE
\8q^n:= \delta q L^n{}_{\0X'}, \quad \8p_n=\delta p L^{{\0Y}}{}_n, \quad \Fi:=\delta r L^{{\0Y}}{}_{\0X'},
\EE

For an alternative representation see  Baugh  \cite{BAUGH2004}.

The space-time simplification introduced 
a large quantum number $l_{\3X}$, 
setting the maximum of the space-time $iL_{\0{XY}}$, and
approaching $\infty$ in the singular limit.
This determines the dimension $N$ of the space-time mode-vector space.
Now the field algebra simplification
 introduces another  large quantum number
$l_{\3F}$ determining the maximum eigenvalue of
the field $L_{\0{X'Y'}}$. 
The simplified relations include
\BE
 [\8q^m, \8p_n]=i \delta q \delta p \left(\delta^n_m L^{\0X'}_{\0Y'}\right) \to i\hbar \delta^n_m
 \EE
 We infer that
 \BE
 l\,\delta q\,\delta p =\hbar
 \EE
 
This simplifies the Lie algebra.  Now we must simplify its representation.
To construct the physical variables, which typically  have many more eigenvalues,  we must pass from the given low-dimensional Lie algebra to a suitable irreducible orthogonal representation of dimension large enough to pass for infinite. 

In the singular theory this representation  
is the bosonic quantification
of the underlying Lie algebra,
unique up to one quantum constant $\hbar$
but infinite-dimensional.
Here in the regular theory 
the space-time Lie algebra is simple,
that of $\o{SO}(5,1)$,
and its representation has finite dimension $\8D_{{\3X}}$,
defined by central invariants $\8{\2c}_{\3X}$.
The representation of the physical basis is defined
by quantum constants $\8{\2h}_{{\3X}}$
close to  their singular values.
Then the simplified boson Lie algebra $\8{\3L}_{\3F}$
has finite dimension 
\BE
\8D_{{\3F}}= 2 \6D_{{\3X}}+2.
\EE
The operational algebra has finite dimension determined
by the central invariants $\8c_{\3F}$ of 
the simplified field Lie algebra.

\subsection{Singular scalar dynamics}

The usual  scalar Green's function is
\BE
\label{eq:GREEN}
G(x'_1, \dots, x'_n)=
 \<\text{vac}| \phi(x'_1), \dots, \phi(x'_n) ) |\text{vac} \>
\EE
Here $x'_1$ is a collection of c numbers, 
eigenvalues of the coordinate operators $x=(x^{\mu})$, 
and $\phi(x'_1)$ is a creation/annihilation 
operator associated with the position 
eigenvalue  $x'_1$.

The construct $G$ is covariant under the unitary group of
basis changes for the space $F$ of fields $\phi(x')$.
Any orthonormal frame $\{\phi_{\alpha}\}$ for the mode-vector space of a single boson defines
a generalized Green's function
\BE
G_{\alpha_1, \dots, \alpha_n}= 
\<\text{vac}|\phi_{\alpha_1}, \dots, \phi_{\alpha_n} |\text{vac} \>
\EE
This form can survive the simplifying that we carry out.
The nature 
of the one-quantum mode-vector, however, changes discontinuously
at the singular limit.
For example, in c space-time the coordinates $x^{\mu}$
all commute,
and so their eigenvalues can label the mode-vector $\phi_{x'}$.
But in the simplified quantum space (STiME),
space-time coordinates $\8x^{\mu}$ do not commute
and their eigenvalues  cannot  label 
a basis.
Instead there are commuting variables 
$t=\delta t\, L_{0 {\0X}}$,
$p_x=\delta p\, L_{1 {\0Y}}$, and $L_{23}$,
which may be supplemented
by  the quantum numbers  $c_2, c_4, c_6$ as necessary 
to make a complete commuting set.
To recover the singular Green's function from the regular
we must construct coherent states that are
only approximately eigenvectors of all the $\8x^{\mu}$\/.

The vacuum mode-vector $|\0{vac}\>$ of the singular quantum theory  is defined
 by its amplitude,
 which has the Lagrangian form
\BE
\<\phi(\cdot)|\text{vac}\>:=N\exp  i \left[ \int d^4 x\, L(\phi(x), \partial_{\mu}\phi(x))\right]
=:N\exp iA\/.
\label{eq:SCALARVACUUM}
\EE
in which $A=A[\phi(.)]$ is the action integral of the exponent.
This gives an amplitude for each field history $\phi(\cdot)$.

The singular dynamical theory we simplify  is that of a free scalar meson, with
Lagrangian density 
\BE
L(\phi(x), \partial_{\mu}\phi(x)):=
-\frac 12 \partial_{\mu}\phi(x) \partial^{\mu}\phi(x)
 +m^2 \phi(x)^2
\EE

\subsection{Regular scalar dynamics}

The  free field or many-quantum action $A$ is 
constructed from the one-quantum antisymmetric operator
\BE
A_1:=i p^{\mu}p_{\mu}+i m^2\/
\EE
by quantification.
To quantify $A_1$, 
we first make explicit the mode-vectors $\phi_x$ and their duals $\phi_x\adj$
that enter into it, and then give  a Lie-algebra on them.
We choose an $x$ basis only for its familiarity:
\BE
A_1= N\int d^4x \, \phi_x L^{xx'}\phi_{x'}\adj\/,
\EE
with a singular normalizer $N$ and 
a singular kernel $L^{x x'}$.
Then quantification
replaces the one-quantum mode-vectors 
$\phi_x$ and their duals $\phi_x\adj$
by many-body operators 
$\iota \phi_x$ and $\phi_x\adj\iota\adj$
obeying  bosonic
commutation relations,
defining the same singular algebra
as a particle in infinite-dimensional space.
The result is 
 the singular action $A$ of 
 (\ref{eq:SCALARVACUUM}),
 now written
\BE
A= N\int d^4x \, \iota \phi_x  L^{xx'}\phi_x\adj\iota\adj\/=
\iota A_1 \iota\adj
\EE
To simplify $A$ we need only simplify $A_1$.

To be sure, the algebra of $\iota$ and $\iota\adj$
is singular and infinite dimensional.
Perhaps it too can be regularized.
This would modify the  q set theory
to allow membership loops,
as in Finsler set theory.
But we do not iterate $\iota$, so it introduces no singularities,
and we leave  $\iota$ fixed.

To simplify the action $A_1$  we simplify each operator in it.
As usual, quantization requires us to order operators that no longer commute
so that their product remains antisymmetric.
For economy we choose the order
\BE
\8A_1=\8p^{\mu}\Fi \8p_{\mu}+m^2\Fi \/.
\EE
The simplified action is then
\BE 
\8A=\8\iota \8A_1 \8\iota\adj
\EE
The simplified creators and annihilators obey simplified bosonic commutation relations, those of 
$d\0{SO}(M)$ with cosmologically large $M$.

Obviously,  this is finite and so is the normalization constant $\8N$ replacing the infinite constant $N$.
The exact Lorentz invariance and the approximate medium-energy Poincar\'e 
invariance are also plausible.

This simplified action seriously breaks the simplified symmetry group, and more symmetric ones that are still good approximations to the singular actgion in the correspondence domain are readily available.
They go beyond the scope of this paper.

\section{Results }

\label{sec:RESULTS}

We have used  general quantization 
 to convert the usual singular theory 
of  the scalar meson  to a finite theory with nearly the
same algebras and symmetries in a correspondence domain.
This toy taught us  how to general-quantize 
Minkowski space-time
and bosonic statistics,
and  how to supply a relativistic finite dynamics
to go with the finite quantum kinematics.

Simplifying space-time quantizes momentum-energy and  space-time.
It produces a finite unified quantum space-time-$i$-momentum-energy space STiME.

The quantization of Minkowski space-time exhibited here has chronons
with simplified bosonic statistics and the symmetry group $\0S\0O(5,1)$\/.
It is a transient theory but some of its features 
are typical.
For one thing,
it  is intrinsically
non-local in both space and momentum variables
with respective non-localities $\delta x$ and $\delta p$.
It also has an invariant integer parameter $\4N$,
a maximum number of elementary processes.
The ether crystallization
breaks Born reciprocity
in the singular limit
$\delta x\to 0$, $\delta p \to \infty$, $\4N\to \infty$\/,
and makes the singular limit theory  local in space-time 
but not in energy-momentum.
That is, in a single interaction
there is no finite change in position or time,
but an arbitrarily large change in momentum and energy;
the standard assumption.

The principle difference between this approach and most others
is that we take seriously the partition 
of the theory into logical levels, each with its algebra,
and preserve these algebras, with 
small changes,
throughout the construction.
This contrasts, for example, with approaches 
to quantum field theory that discretize
space-time, discarding the Lorentz invariance, and then take a limit.
Under general quantization the
system determines its own quanta
and requires no ad hoc discretization.

The correspondence principle fixes some combinations 
of  the new quantum constants, quantum numbers,
and regulants, leaving the rest to experiment.
No infinite renormalization is needed.

Several discrete choices  have to be left to experiment.
For example the simplicity principle
is equally satisfied along the real, complex, and quaternionic
lines of simple Lie algebras.
We chose the real line mainly because it is easiest
and  in some sense simplest,
but nature may not take the way that is easiest or simplest for us.

We give necessary conditions
on the defining parameters
for the finite theory to converge to the usual theory
in some appropriately weak sense,
but we have not shown they are sufficient.
This question may be sensitive to the theory under study.
We have not proven 
 that these finite results agree well enough with the finite results of the usual singular theory where they should
 but it seems plausible.
Approximating 
the regular discrete spectrum by a singular continuous one is a somewhat delicate 
non-uniform convergence even for the harmonic oscillator.

Such a change in the most basic algebraic relation of quantum theory
and in the Heisenberg Uncertainty Relation has many experimental consequences to be developed.
Vilela Mendes points out that it permits a serious reduction in phase space  $\Delta p \,\Delta q$ at high energy that may explain the GZK anomaly
\cite{VILELA2005}. 

We suspend our study of the scalar field for now in order
 to general-quantize more basic systems, the
gauge fields mediating
physical interactions.

\section{Acknowledgments}

Helpful discussions are gratefully acknowledged
with James Baugh, Eric Carlen,  
Giuseppe Castagnoli,
Andrei Galiautdinov, Viqar Husain, Alex Kuzmich, 
Zbigniew  Oziewicz, 
Frank Schroek, Mohsen Shiri-Garakani,
Heinrich Saller, Raphael Sorkin, and John Wood.
For  hospitality and support during some of this work,
I thank the Heisenberg Institute of Theoretical Physics, Munich,
the Elsag Corporation, Genoa,
the Library of Tibetan Works and Archives,
and the Mathematical Research Institute of Oberwolfach.

\newpage


\begin{thebibliography}{}

\bibitem{ANDERSON} Anderson, P. W. (1972) 
More is different. {\em Science}
{\em 177}, 393-396. 


\bibitem{Atakishiyev2003} Atakishiyev, N. M., G. S. Pogosyan, and K. B.
Wolf, Contraction of the finite one-dimensional oscillator, {\it
International Journal of Modern Physics} {\bf A18}, 317, 2003.


\bibitem{BACON} Bacon, F. {\it Novum Organum} (1620). P. Urban and J.
Gibson (tr. eds.) Peru, Illinois: Open Court Publishing Company,
1994.

\bibitem{BAUGH2004} Baugh, J. (2004)  Regular Quantum Dynamics. Ph. D. Thesis, School of Physics, Georgia Institute of Technology.

\bibitem{BERGMANN1972} Bergmann, P. G. and A. Komar (1972). 
{\em International Journal of Theoretical Physics} {\bf 5}, 15.

\bibitem{BERGMANN1979} Bergmann, P. G. (1979) The Fading World Point. 
In  
Bergmann, P. G. and V. de Sabbata  (editors),
{\em Cosmology and Gravitation. Spin, Torsion, Rotation, and Supergravity}\/.
Pages  173-176.
Plenum Publishing Co. 



\bibitem{BORN} 
Born, M., K. C. Cheng, and H. S. Green (1949) {\em Nature} {\bf 164}: 281-282.
Reciprocity theory of electrodynamics. 
Born, M. (1949) {\em Reviews of Modern Physics} {\bf 21}: 463-473.
Reciprocity theory of elementary particles

\bibitem{Carlen2001}Carlen, E. and R. Vilela Mendes (2001).
Non-commutative space- time and the
uncertainty principle. {\it arXiv: quant-ph/0106069}.
 {\em Physics Letters} {\bf A 290}:109-114

\bibitem{CONNES} Connes, A. (1994) {\em Non-commutative geometry}\/.  
	Academic Press, San Diego, CA.

\bibitem{CZACHOR} Czachor, M. and M. Wilczewski (2005) 
Direct test of representation of canonical commutation relations
employed in field quantization 
quant-ph/0509117

\bibitem{FIALOWSKI} Fialowski, A.  (1986) Deformations of Lie algebras. {\em Math. USSR Sbornik}  {\bf 35}: 467-473


%
%


\bibitem{FINKELSTEIN1969} Finkelstein, D. (1961) Space-time code, {\it
Physical Review } {\bf 184}, 1261. 
Space-time code II, {\it Physical Review}\/  D5, 320- 
(1972). Space-time code III, {\it Physical Review}\/  D5,  2922-
(1972)
Space-time code IV.
{\it Physical Review}\/  D9, 2219-
(1974)

\bibitem{FINKELSTEIN1974}
D. Finkelstein, G. Frye, and L. Susskind, Space-time code V, {\it Physical Review}\/  D9,  2231 (1974)

\bibitem{FINKELSTEIN1982} Finkelstein, D.  Quantum set theory and Clifford algebra, {\it International Journal of Theoretical Physics}\/  21, 489-503 (1982)

\bibitem{FINKELSTEIN1996} Finkelstein, D. R. {\it Quantum Relativity}.
Heidelberg: Springer, 1996.



%
\bibitem{FLATO} Flato, M. (1982) Deformation view of physical theories.
{\em Czechoslovak Journal of Physics} {\bf B32}, 472-475.

\bibitem{GALIAUTDINOV2002}
Galiautdinov A.A., and D. R. Finkelstein (2002). Chronon corrections to the
Dirac equation, {\it Journal of Mathematical Physics} {\bf 43}, 4741
(2002).

\bibitem{GEROCH} Geroch, R. (1972)  Einstein Algebras. {\em Communications of Mathematical Physics} {\bf 26}, 271-275. 

\bibitem{GERSTENHABER1964} Gerstenhhaber, M. (1964) 
{\em Annals of Mathematics} {\bf 32}, 472


\bibitem{GOLDEN} Golden, G. A. and S. Sarkar. Local currents for a deformed algebra of quantum mechanics with a fundamental scale.
Preprint, Jan. 13, 2006

\bibitem{INONU1952} In\"on\"u, E. and Wigner, E.P. {\it Proceedings of the  National 
Academy of  Science} {\bf 39}, 510-524, 1953.

\bibitem{INONU1964} In\"on\"u, E. in F. G\"ursey, (ed.){\it Group Theoretical Concepts
and Methods in Elementary Particle Physics.}  New York: Gordon \&
Breach, 391-402, 1964.

\bibitem{KIM1987}  Kim , Y. S. and E. P. Wigner.  Cylindrical group and massless  particles.  Ê{\em Journal of Mathematical Physics} 28, pages 1175-1179 (1987).
\bibitem{KIM1990}   Kim , Y. S. and E. P. Wigner.   Space-time geometry of relativistic  particles. Ê{\em Journal of Mathematical Physics} 31, pages 55-60 (1990).

\bibitem{KUZMICH} Kuzmich, A., N. P. Bigelow, and L. Mandel, 
{\it Europhysics Letters } {\bf A 42}, 481 (1998).
Kuzmich, A., L. Mandel, J. Janis, Y. E. Young, R.
Ejnisman, and N. P. Bigelow, {\it Physical  Review A} {\bf 60}, 2346 (1999).
Kuzmich, A., L. Mandel, and N. P. Bigelow, {\it Physical Review Letters} 
{\bf 85}, 1594 (2000).

\bibitem{LAUGHLIN2000} Laughlin, R. D. and D. Pines (2000),
The Theory of Everything. {\it
Proceedings of the National Academy of Sciences} {\bf 97}, 28 

\bibitem{MADORE1992} Madore, J. (1992) {\em Classical and Quantum Gravity} {\bf  9}, 69-87.
 The fuzzy sphere
\bibitem{MADORE1999}  Madore, John and N. J. Hitchin (1999)
{\em An Introduction to Noncommutative Differential Geometry and its Physical Applications} 2nd edition. Cambridge University Press 

\bibitem{NEUMANN1931} von Neumann, J. (1931) 
Die Eindeutigkeit der Schršdingerschen Operatoren.
{\em Mathematische Annalen} {\bf 104}, 570-578. 

\bibitem{NEUMANN1932} von Neumann, J. (1932) 
{\em Mathematische Grundlagen der Quantenmachanik}\/. Springer.







\bibitem{SALLER2006} Saller, H. (2006) {\em Operational Physics}\/. Springer.


\bibitem{SEGAL} Segal, I. E. (1951) {\it Duke Mathematics Journal} {\bf 18}, 221-265.
Especially pp. 255-256.

\bibitem{SHIRI2004} Shiri-Grakani M. Finite quantum theory of the 
harmonic oscillator.  Ph. D. Thesis, School of Physics, Georgia Institute of Technology, 2004.

\bibitem{SHIRI2005} Shiri-Garakani, M. and D. R. Finkelstein
Finite Quantum Theory of the Harmonic Oscillator.
quant-ph/0411203.
Based on \cite{SHIRI2004}.

\bibitem{SNYDER} Snyder, H.S. (1947) {\it Physical Review } 71: 38-41.

\bibitem{THOOFT} 'tHooft, G.  (2003) Determinism in Free Bosons.
{\em International Journal of Theoretical Physics} 42: 355.

\bibitem{VILELA1994}  Vilela Mendes, R.  (1994)  {\it Journal of Physics A: Math. Gen.}
 27: 8091-8104.

\bibitem{VILELA1996}  Vilela Mendes, R. (1996)
{\it Physics Letters}
 A210: 232.

\bibitem{VILELA2000}  Vilela Mendes, R. (2000)
{\it J. Math. Phys.}
41: 156.

\bibitem{VILELA2005}  
  Vilela Mendes, R.  (2005) Some consequences of a noncommutative space-time structure.
  hep-th/0406013.
  {\em European Physics Journal} C 42: 445-452 
  
  \bibitem{WEIZSACKER}von~Weizs\"acker, C. F. (1955).
Komplementarit\"at und logik.
{\em Naturwissenschaften}, 42:521-529, 545-555.

\bibitem{WEIMAR} Weimar-Woods, E.  (1995)  {\em Journal of  Mathematical  Physics} 36: 4519

\bibitem{WESS2002} Wess, J.. Non-Abelian gauge theories on non-commutative spaces. http://www-library.desy.de/preparch/desy/proc/proc02-02/Proceedings/pl.7/wess\_pr.pdf.


\bibitem{WESS2004} Wess, J. Deformed coordinate space: derivatives. hep-th/0408080


  
\bibitem{WILCZEK1982}   Wilczek, F.  (1982). {\it Physical Review  Letters} {\bf  48}, 1144.

\bibitem{WILSON}
Wilson,ÊK.ÊG. (1974) Confinement of quarks
{\em Physical Review} {\bf D10}, 2445-2459
\end{thebibliography}
\end{document}